# Effect of spin relaxation rate on the interfacial spin depolarization in ferromagnet/oxide/semiconductor contacts

Kun-Rok Jeon[1], Byoung-Chul Min[2], Youn-Ho Park[2], Young-Hun Jo[3], Seung-Young Park[3], Chang-Yup Park[1], and Sung-Chul Shin[1]

[1]*Department of Physics and Center for Nanospinics of Spintronic Materials, Korea Advanced Institute of Science and Technology (KAIST), Daejeon 305-701, Korea*

[2]*Center for Spintronics Research, Korea Institute of Science and Technology (KIST), Seoul 136-791, Korea*

[3]*Nano Materials Research Team, Korea Basic Science Institute (KBSI), Daejeon 305-764, Korea*

## Abstract

Combined measurements of normal and inverted Hanle effects in CoFe/MgO/semiconductor (SC) contacts reveal the effect of spin relaxation rate on the interfacial spin depolarization (ISD) from local magnetic fields. Despite the similar ferromagnetic electrode and interfacial roughness in both CoFe/MgO/Si and CoFe/MgO/Ge contacts, we have observed clearly different features of the ISD depending on the host SC. The precession and relaxation of spins in different SCs exposed to the local fields from more or less the same ferromagnets give rise to a notably different ratio of the inverted Hanle signal to the normal one. At room temperature, a large ISD is observed in the CoFe/MgO/Si contact, but a small ISD in the

CoFe/MgO/Ge contact. The ISD of the CoFe/MgO/Ge contact has been substantially increased at low temperature. These results can be ascribed to the difference of spin relaxation in host SCs. A model calculation of the ISD, considering the spin precession due to the local field and the spin relaxation in the host SC, explains the temperature and bias dependence of the ISD consistently.

**1. Introduction**

The electrical injection and detection of spin-polarized carriers in semiconductors (SCs) has been successfully achieved by employing spin tunnel contacts [1-15]. However, many aspects of the spin phenomena in these systems [1-13,16], e.g., (i) the location, magnitude, and sign of the induced spin accumulation, (ii) the unusual bias and temperature-dependence of the spin signal, and (iii) the unexpected short spin lifetime and its weak variation with temperature, require additional investigation.

Recently, Dash *et al.* have shown the effect of local-field strength on the spin signals in ferromagnet (FM)/oxide/SC contacts; it was found that the local magnetostatic fields ($B_L^{ms}$) arising from the finite roughness of the FM/oxide interface dramatically alter and even dominate the accumulation and dynamics of the spins in SCs [17]. Because this interfacial spin depolarization (ISD) due to $B_L^{ms}$ is deeply interconnected with (i), (ii), and (iii) [17], a systematic study of the ISD is crucial for a complete understanding of the spin accumulation and spin dynamics in SC near FM interface. The inverted Hanle effect due to local-fields had been extensively studied in Ref 17, using the FM/Al$_2$O$_3$/Si contacts with the same host SC but different FMs [17]. FM/Al$_2$O$_3$/GaAs contacts also showed a similar signature of the ISD with slightly different details, suggesting that the ISD is universal for the three-terminal Hanle (TTH) experiments [17].

In this vein, it is of interest to investigate the role of host SCs on the ISD in FM/oxide/SC contacts. Here we report, using the FM/oxide/SC contacts with the same FM but different host SCs, the effect of spin relaxation rate on the ISD. The combined measurements of normal and inverted Hanle effects over wide temperature ($T$) and bias current ($I$) range reveal the effect of spin relaxation rate on the ISD in CoFe/MgO/Si and CoFe/MgO/Ge contacts. We have observed, despite more or less the same ferromagnetic electrode and the interfacial roughness of the FM/oxide/SC contacts, significant differences of the ISD depending on the host SC; the spin accumulation in different SCs exposed to the local fields from similar ferromagnets gives rise to a clearly different ratio of the inverted Hanle signal to the normal one. This can be understood in terms of two competing mechanisms in the host SCs, namely the spin relaxation and spin precession due to the local fields.

## 2. Experimental details

Two types of CoFe(5 nm)/MgO(2 nm)/$n$-SC(001) tunnel contacts were prepared using a molecular beam epitaxy system. The first one is a highly ordered CoFe/MgO/Si contact where the Si channel is a heavily As-doped ($n_d \sim 2.5 \times 10^{19}$ cm$^{-3}$ at 300 K) [6], and the second one is a single-crystalline CoFe/MgO/Ge contact where the Ge channel consists of a heavily P-doped surface layer ($n_d \sim 10^{19}$ cm$^{-3}$ at 300K) and a moderately Sb-doped substrate ($n_d \sim 10^{18}$ cm$^{-3}$ at 300K) [10]. In order to measure the induced spin accumulation ($\Delta\mu = \mu^\uparrow - \mu^\downarrow$) in the spin tunnel contacts, we have fabricated devices for the TTH measurements [1-4], consisting of multiple CoFe/MgO/$n$-SC tunnel contacts

($100 \times 100$ $\mu m^2$). Details of the sample preparation as well as the structural and electrical characterization are available in the literature [6,10,18]. It should be noted that the dominant transport mechanism for both CoFe/MgO/Si and CoFe/MgO/Ge contacts is tunneling, as proven by the symmetric *I-V* curve and its weak temperature-dependence; the both types of contacts reveal the narrow depletion region of ~5 nm and the small resistance-area values ($\sim 5 \times 10^{-6}$ $\Omega\,m^2$ (300 K) to $\sim 1 \times 10^{-5}$ $\Omega\,m^2$ (5 K) at -0.25 V) [6,10].

## 3. Results & discussion

### 3.1. Roughness and magnetization characterization of tunnel contacts

To estimate the magnitude of $B_L^{ms}$, which scales with the roughness of the FM interface and the magnetization of the FM [17], we have characterized the roughness of an MgO/SC reference structure without FM using atomic force microscopy (AFM), and magnetic property of complete FM/MgO/SC structures using vibrating sample magnetometry (VSM). The MgO/SC reference structure show a root-mean-square (RMS) roughness of ~0.2 nm, peak-to-peak height variations of ~0.3-0.4 nm, and lateral correlation lengths of 30 to 50 nm. The FM/MgO/SC samples have saturation magnetization ($M_s$) values of ~1400-1430 emu/cc with a normalized remanence ($M_r/M_s$, where $M_r$ is remanent magnetization) of ~0.93-0.95 for the easy axis magnetization. Because the depletion region width of both CoFe/MgO/Si and CoFe/MgO/Ge contacts are more or less the same (about 5 nm) [6,10], the locations of spin accumulation in both contacts are likely to be similar to each other. Taking into account the similar roughness, magnetization, and depletion width, it is likely that the magnitude of $B_L^{ms}$ at the interface is not fundamentally different in both CoFe/MgO/Si and CoFe/MgO/Ge contacts.

## 3.2. High-field Hanle measurements of tunnel contacts and their generic features

Using the CoFe/MgO/Si and CoFe/MgO/Ge contacts, we have conducted high-field TTH measurements (up to ± 3 T) under perpendicular ($M \perp B$, closed circles) and in-plane ($M//B$, open circles) magnetic field (figure 1). The TTH measurements show the overall behavior of the spin accumulation signals as a function of external magnetic fields ($B_\perp$, $B_\parallel$).

As indicated in figure 1, three distinct regions were observed in the $M \perp B$ measurements (closed circles): (*i*) the Hanle effect at small magnetic fields, (*ii*) the rotation of magnetization (*M*), and (*iii*) the saturation of *M*. As $B_\perp$ is increased, the voltage signal from the spin accumulation is sharply reduced due to the Hanle effect in region (*i*), and thereafter it gradually increases when the *M* of the FM rotates out of the plane in region (*ii*). When the *M* and induced spin accumulation in the SC are fully aligned with $B_\perp$ higher than the demagnetization field (~2.2 T) of CoFe, the voltage signal eventually becomes saturated in region (*iii*). Both CoFe/MgO/Si and CoFe/MgO/Ge contacts show a very similar field dependence of spin signals in the $M \perp B$ measurements.

On the other hand, the *M//B* measurement (open circles) shows a clear difference in the field dependence of spin signals between CoFe/MgO/Si and CoFe/MgO/Ge contacts. We have observed a sizable inverted Hanle effect [17] in the CoFe/MgO/Si contact. At zero or small external magnetic fields, the injected spins are precessed and dephased by local magnetostatic fields having random directions, which result in a reduction of the spin accumulation [17]. In contrast, a larger external magnetic field ($B_\parallel$) can eliminate the local magnetostatic fields, restoring a full spin accumulation [17]. The overall behavior of the voltage signals (for the perpendicular and in-plane field) is in good agreement with the findings of the previous study on FM/Al$_2$O$_3$/SC tunnel contacts [17].

In the *M//B* measurement, the CoFe/MgO/Ge contact shows two clear differences from the CoFe/MgO/Si contact. First, the spin signals in the saturation region (*iii*) are significantly different in the *M*⊥*B* and *M//B* measurements. This is a consequence of the tunneling anisotropic magnetoresistance (TAMR) [19-23]; the tunnel resistance depends on the angle between the *M* and crystal axes of FM because the tunneling electrons experience the anisotropic density of states with respect to *M* via the spin-orbit interaction. The effect of TAMR on the spin signals has been also observed in the Fe/MgO/Ge and Co/Al$_2$O$_3$/GaAs tunnel contacts [8,17]. The positive background signal in the CoFe/MgO/Ge contact may be ascribed to the Lorentz MR (LMR) in Ge substrate due to its high mobility (note that the LMR is quadratic in mobility). The second difference between two contacts is that the CoFe/MgO/Ge contact has a small magnitude of the inverted Hanle effect. Considering that the injected spins experience a similar magnitude of $B_L^{ms}$, the difference in the magnitude of the in-plane Hanle signal is rather unexpected.

### 3.3. Competition between spin relaxation and spin precession

The spin injection under the random $B_L^{ms}$ result in the precession of injected spins with an angular frequency of $\omega_L^{ms} = g\mu_B B_L^{ms}/\hbar$, where *g* is the Landé *g*-factor, $\mu_B$ is the Bohr magneton, and $\hbar$ is the Planck constant divided by $2\pi$. Accompanying the spin precession in random orientation, the spin relaxation with a spin lifetime of $\tau_{sf}$ takes place as a consequence of the microscopic spin scatterings inside the SC. If $\tau_{sf} \ll 1/\omega_L^{ms}$, the spins are completely relaxed within their spin lifetime before being precessed by $B_L^{ms}$. In this case, the suppression of spin accumulation by $B_L^{ms}$ is negligible, leading to a small $|\Delta V_{inverted}/\Delta V_{normal}|$. In contrast, if $\tau_{sf} \geq 1/\omega_L^{ms}$, the $|\Delta V_{inverted}/\Delta V_{normal}|$ becomes pronounced because the spins are precessed many times in $B_L^{ms}$ and randomized within their $\tau_{sf}$, resulting in the sizable suppression of the spin polarization by $B_L^{ms}$.

Here we define $R_{ISD} \equiv |\Delta V_{inverted}/\Delta V_{normal}|$, the ratio of the inverted Hanle signal to the normal one, which is quite a good measure of the ISD in CoFe/MgO/SC contacts. Despite the similar magnitude of $B_L^{ms}$, the inverted Hanle signals in figure 1 clearly show that the $R_{ISD}$ is more pronounced for the CoFe/MgO/Si contact than it is for the CoFe/MgO/Ge contact at 300 K.

The effective strength of $B_L^{ms}$ for spins located 7 nm away from the FM interface is about 0.1~1 kOe [17], corresponding to a $1/\omega_L^{ms}$ value of about 0.09~0.9 ns for Si ($g = 2$) and 0.11~1.1 ns for Ge ($g = 1.6$). The $\tau_{sf}$ estimated from the spin scattering via the Elliot-Yafet (EY) mechanism [24-26] is about 1 ns for heavily-doped Si at 300 K. Because this $\tau_{sf}$ is comparable to $1/\omega_L^{ms}$, a large $R_{ISD}$ can be expected for the CoFe/MgO/Si contact.

If the $\tau_{sf}$ in Ge at 300 K is small due to the non-negligible spin-orbit interaction or other scattering mechanisms, the relatively weak $R_{ISD}$ in the CoFe/MgO/Ge contact at 300 K can be explained.

## 3.4. Strong enhancement of interfacial spin depolarization at low temperatures

The interpretation above appears even more convincing given the strong enhancement of the $R_{ISD}$ with temperature ($T$). Figure 2 shows the normal ($\Delta V_{normal}$) and inverted ($\Delta V_{inverted}$) Hanle effects on the CoFe/MgO/Si and CoFe/MgO/Ge contacts under $M \perp B$ (closed circles) and $M // B$ (open circles) measurements, respectively, at an applied current ($I$) of -500 µA (spin injection condition) with various temperatures. All curves are normalized by the voltage difference between the minimum value of the normal Hanle curve and the saturation value of the inverted Hanle curve in the region *(i)* (see figure 1). The magnitude of the inverted Hanle effect of the CoFe/MgO/Ge contact has been measured excluding the background LMR effect.

From the Lorentzian fit of the Hanle curves, we can determine the effective $\tau_{sf}$ ($\tau_{eff}$) of accumulated spins. It is difficult to extract the true or real $\tau_{sf}$ from the Hanle curve (with the Lorentzian fit) due to the artificial broadening caused by the ISD. Nevertheless, the effective value of $\tau_{sf}$ (or $\tau_{eff}$), which should be considered as a lower bound for $\tau_{sf}$, can be deduced. For a quantitative comparison, we have plotted the $R_{ISD}$ and $\tau_{eff}$ as a function of $T$ in figure 3.

As $T$ decreases from 300 K to 5 K, the $R_{ISD}$ becomes larger and the $\tau_{eff}$ increases gradually; the $R_{ISD}$ for the CoFe/MgO/Si contact is continually enhanced ~2 times (figure 3(a)); the increase of $R_{ISD}$ for the CoFe/MgO/Ge contact is even more pronounced; the $R_{ISD}$ is increased ~9 times. Taking into account that $B_L^{ms}(T) \propto (1-\alpha T^{3/2})$ with $\alpha = 3.2 \times 10^{-5} K^{-3/2}$ for a CoFe [27,28], the $\omega_L^{ms}$ is slightly increased with decreasing $T$. It is clear that the large enhancement of the $R_{ISD}$ at low $T$ is mainly originated from the increase of $\tau_{sf}$, as expected in the EY mechanism [24-26].

The temperature dependence of $R_{ISD}$ corresponding to the $\tau_{eff}$ variation supports the interpretation based on two competing mechanisms in the host SCs, namely the spin relaxation and spin precession due to the local fields.

### 3.5. Theoretical description of the effect of spin relaxation on interfacial spin depolarization

Using the model in Ref. 17, we have calculated the effect of $\tau_{sf}$ on the Hanle curves with a fixed value of $1/\omega_L^{ms}$. According to the model [17], the $S_x$ component of steady state spin polarization $\vec{S}$ at the interface, which is parallel to the $M$ vector of FM detector, in the presence of local magnetostatic field ($B_L^{ms}$) and external applied magnetic field ($B^{ext}$) is expressed as:

$$S_x = S_{0x}\left\{\frac{\omega_x^2}{\omega_L^2} + \left(\frac{\omega_y^2 + \omega_z^2}{\omega_L^2}\right)\left(\frac{1}{1+(\omega_L \tau_{sf})^2}\right)\right\} \quad (1)$$

where $S_{0x}$ is the spin polarization without any magnetic field, $\omega_L^2 = \omega_x^2 + \omega_y^2 + \omega_z^2$, and

$\omega_i = \omega_i^{ext} + \omega_i^{ms}(x,y,z)$. Here, $\omega_i^{ms}(x,y,z)$ was taken to have a periodic spatial variation with $\omega_L^{ms} \cos(2\pi x/\lambda)$, where $\omega_L^{ms} \approx 3$ ns$^{-1}$ (or $1/\omega_L^{ms} \approx 0.33$ ns, corresponding to a $B_L^{ms}$ value of 0.3 kOe) and $\lambda = 40$ nm and where the spin polarization was averaged in space over a full period $\lambda$ for simplicity.

Figure 4 shows the calculated normal ($M \perp B$) and inverted ($M // B$) Hanle curves, which qualitatively reproduce the experimental results. As $\tau_{sf}$ is increased (with a fixed value of $1/\omega_L^{ms} \approx 0.33$ ns), the inverted Hanle effect (brown symbols) becomes pronounced; the $R_{ISD}$ has been also increased; the widths of the normal Hanle curve (wine symbols) and inverted Hanle curve are broadened in comparison with the ideal Hanle curve (blue symbols) without $B_L^{ms}$.

We have also calculated the $R_{ISD}$ as a function of $\tau_{sf}$ with the four different $1/\omega_L^{ms}$ values of 0.10, 0.33, 1.00, and $\infty$ ns (corresponding to $B_L^{ms}$ values of about 1.0, 0.3, 0.1, and 0.0 kOe, respectively):

$$R_{ISD} \equiv \frac{S_{0x} - S_x(B_{ext}=0)}{S_x(B_{ext}=0)} = \frac{\left(\left(\omega_y^{ms}\right)^2 + \left(\omega_z^{ms}\right)^2\right)\left(\omega_L^{ms}\tau_{sf}\right)^2}{\left(\omega_x^{ms}\right)^2\left(1+\left(\omega_L^{ms}\tau_{sf}\right)^2\right) + \left(\omega_y^{ms}\right)^2 + \left(\omega_z^{ms}\right)^2} \quad (2)$$

where $\left(\omega_L^{ms}\right)^2 = \left(\omega_x^{ms}\right)^2 + \left(\omega_y^{ms}\right)^2 + \left(\omega_z^{ms}\right)^2$. When $\tau_{sf} \gg 1/\omega_L^{ms}$, the $R_{ISD}$ is determined only by the ratio of each component, $\left(\left(B_y^{ms}\right)^2 + \left(B_z^{ms}\right)^2\right)/\left(B_x^{ms}\right)^2$ or $\left(\left(\omega_y^{ms}\right)^2 + \left(\omega_z^{ms}\right)^2\right)/\left(\omega_x^{ms}\right)^2$.

As depicted in figure 4(d), the $R_{ISD}$ is strongly enhanced at high values of $\tau_{sf}$; when $1/\omega_L^{ms}$ is small, the $R_{ISD}$ increases more rapidly as a function of $\tau_{sf}$. The calculated $R_{ISD}$ are relatively smaller than the observed values because the magnitudes of three components $(B_x^{ms}, B_y^{ms}, B_z^{ms})$ are assumed to be the same for simplicity.

Two important points can be obtained from the $\tau_{eff}$ vs. $\tau_{sf}$ plot with different values of $1/\omega_L^{ms}$ in figure 4(e) (note that the $\tau_{sf}$ is the input value for the calculation and the $\tau_{eff}$ is the extracted value from the calculated normal Hanle curve, see figures. 4(a)-(c)). The first point is that the $\tau_{eff}$ is significantly dependent on the $1/\omega_L^{ms}$. For example, for the small values of $1/\omega_L^{ms}$, the $\tau_{eff}$ significantly deviates from the $\tau_{sf}$. The second one is that, in spite of the artificial broadening by $B_L^{ms}$, the $\tau_{eff}$ still is a monotonically increasing function of the $\tau_{sf}$. The increase of the $\tau_{sf}$ results in the increase of both $\tau_{eff}$ and $R_{ISD}$. This agrees with our experimental finding that the $R_{ISD}$ is positively correlated with the $\tau_{eff}$.

Based on the reasonable agreement of the calculated curves with observed ones, we conclude that his model basically explains the experimental finding quite well and captures the basic physics.

### 3.6. Decrease of interfacial spin depolarization under higher reverse bias

The analysis on the bias dependence of the $R_{ISD}$ provides another important evidence to show the influence of spin relaxation rate on the ISD. For high reverse bias, the injected spin-polarized electrons have a high kinetic energy relative to the Fermi-level ($E_{F, SC}$) of the SC, and releases the energy via the thermalization process [29]. It can happen that, during the thermalization process, these electrons lose their spin orientation more easily than the electrons injected at lower energy level. If this is the case, the spin scattering and depolarization becomes stronger under a higher reverse bias, resulting in the decrease of $\tau_{sf}$. A small $\tau_{sf}$ (or a large relaxation rate, $1/\tau_{sf}$) leads to the decrease of the $R_{ISD}$ as explained in the previous section. (note that the broadening of the depletion region with bias current is not significant because the electronic transport for the both contacts is dominated by the MgO tunnel barrier not by the Schottky barrier [6,10]).

We have checked if the bias dependences of the $R_{ISD}$ and $\tau_{eff}$ show a consistent behavior. Figure 5 shows the normal ($\Delta V_{normal}$) and inverted ($\Delta V_{inverted}$) Hanle effects on the CoFe/MgO/Si and CoFe/MgO/Ge contacts under $M \perp B$ (closed circles) and $M // B$ (open circles) measurements, respectively, with various reverse bias currents ($I$ <0, spin injection condition) at 5 K. Notably, as $I$ increases (or much hotter spin-polarized electrons are injected into the SCs), both contacts show a clear and gradual decrease of the $R_{ISD}$ and the broadening of the Hanle curves. Figure 6 summarizes the $R_{ISD}$ and the $\tau_{eff}$ as a function of $I$ at 5 K. In these figures, we can clearly see that the $R_{ISD}$ is proportional to the $\tau_{eff}$, and the $R_{ISD}$ difference between the CoFe/MgO/Si and CoFe/MgO/Ge contacts is originated from the different $\tau_{eff}$.

The result again confirms that the two competing mechanisms, spin precession due to the local fields and the spin relaxation in the host SCs, contribute to determine the shape and magnitude of the normal ($M \perp B$) and inverted ($M // B$) Hanle curves.

## 4. Conclusions

We have investigated the effect of spin lifetime on the interfacial spin depolarization from the local fields in CoFe/MgO/Si and CoFe/MgO/Ge contacts using the combined measurements of normal and inverted Hanle effects. Although both contacts have a similar interfacial roughness and local magnetic field strength, the observed Hanle curves are quite different. The $|\Delta V_{inverted}/\Delta V_{normal}|$ of CoFe/MgO/Si contact is larger than that of the CoFe/MgO/Ge contact at room temperature, and the $|\Delta V_{inverted}/\Delta V_{normal}|$ of the CoFe/MgO/Ge contact shows a strong temperature dependence. These results are associated with two competing mechanisms, spin precession due to the local fields and the spin relaxation in the host SCs and their temperature dependences. The model calculation, considering two competing mechanisms, reproduces the experimental observations quite well. The bias dependences of the Hanle curves also show consistent behaviors.

## Acknowledgments

This work was supported by the National Research Laboratory Program Contract No. R0A-2007-000-20026-0 through the National Research Foundation of Korea funded by the Ministry of Education, Science and Technology, the KIST institutional program, and by KBSI grant number T31405 for Young-Hun Jo.

## References

[1] Lou X, Adelmann C, Furis M, Crooker S A, Palmstrøm C J and Crowell P A 2006 *Phys. Rev. Lett*. **96** 176603

[2] Dash S P, Sharma S, Patel R S, de Jong M P,and Jansen R 2009 *Nature* **462** 491

[3] Jansen R, Min B C, Dash S P, Sharma S, Kioseoglou G, Hanbicki A T, van't Erve O M J, Thompson P E and Jonker B T 2010 *Phys. Rev.* B **82** 241305(R)

[4] Li C H, van 't Erve O M J and B. T. Jonker 2011 *Nat. Commun.* **2** 245

[5] Gray N W and Tiwari A 2011 *Appl. Phys. Lett.* **98** 102112

[6] Jeon K R, Min B C, Shin I J, Park C Y, Lee H S, Jo Y H and Shin S C 2011 *Appl. Phys. Lett*. **98** 262102

[7] Ando Y, Kasahara K, Yamane K, Baba Y, Maeda Y, Hoshi Y, Sawano K, Miyao M and Hamaya K 2011 *Appl. Phys. Lett.* **99** 012113

[8] Saito H, Watanabe S, Mineno Y, Sharma S, Jansen R, Yuasa S and Ando K 2011 *Solid State Comm.* **151** 1159

[9] Jeon K R, Min B.C, Jo Y H, Lee H S, Shin I J, Park C Y, Park S Y and Shin S C 2011 *Phys. Rev.* B **84** 165315

[10] Jeon K R, Min B C, Park Y H, Lee H S, Park C Y, Jo Y H and Shin S C 2011 *Appl. Phys. Lett.* **99** 162106

[11] Jain A, Louahadj L, Peiro J, Le Breton J C, Vergnaud C, Barski A, Beigné C, Notin L, Marty A, Baltz V, Auffret S, Augendre E, Jaffrès H, George J M and Jamet M 2011 *Appl. Phys. Lett.* **99** 162102

[12] Kasahara K, Baba Y, Yamane K, Ando Y, Yamada S, Hoshi Y, Sawano K, Miyao M and K. Hamaya arXiv:1105.1012v3

[13] Hanbicki A T, Cheng S F, Goswami R, van `t Erve O M J and Jonker B T 2012 *Solid State Comm.* **152** 244

[14] Johnson M and Silsbee R H 1985 *Phys. Rev. Lett.* **55** 1790

[15] Johnson M and Silsbee R H 1988 *Phys. Rev.* B **37** 5326

[16] Tran M, Jaffrès H, Deranlot C, George J M, Fert A, Miard A and Lemaître A 2009

*Phys. Rev. Lett.* **102** 036601

[17] Dash S P, Sharma S, Le Breton J C, Jaffrès H, Peiro J, George J M, Lemaître A and Jansen R 2011 *Phys. Rev.* B **84** 054410

[18] Jeon K R, Park C Y and Shin S C 2010 *Cryst. Growth Des*. **10** 1346

[19] Gould C, Ruster C, Jungwirth T, Girgis E, Schott G M, Giraud R, Brunner K, Schmidt G and Molenkamp L W 2004 *Phys. Rev. Lett.* **93** 117203

[20] Saito H, Yuasa S and Ando K 2005 Phys. Rev. Lett. **95** 086604

[21] Gao L, Jiang X, Yang S H, Burton J D, Tsymbal E Y and Parkin S S P 2007 *Phys. Rev. Lett.* **99** 226602

[22] Umemura T, Harada M, Matsuda K and Yamamoto M 2010 *Appl. Phys. Lett*. **96** 252106

[23] Liu R S, Michalak L, Canali C M, Samuelson L and Pettersson H 2008 *Nano Lett*. **8** 848

[24] Elliott R J 1954 *Phys. Rev.* **96** 266 Yafet Y 1963 *Solid State Physics* ed F Seitz and D Turnbull vol 14 (Academic: New York) p 1

[25] Restrepo O D and Windl W *arXiv:1010.5436v2*

[26] Cheng J L, Wu M W and Fabian J 2010 *Phys. Rev. Lett.* **104** 016601

[27] Kipferl W, Dumm M, Rahm M and Bayreuther G 2003 *J. Appl. Phys*. **93** 7601

[28] Wang W, Sukegawa H and Inomata K 2010 *Phys. Rev.* B **82** 092402

[29] Motsnyi V F, Van Dorpe P, Van Roy W, Goovaerts E, Safarov V I, Borghs G and De Boeck J 2003 *Phys. Rev.* B **68** 245319

**Figure captions**

Figure. 1. High-field Hanle measurements (up to ± 3 T) on (a) CoFe/MgO/Si and (b) the

CoFe/MgO/Ge contacts at 300 K under perpendicular ($M \perp B$, closed circles) and in-plane ($M//B$, open circles) measurement schemes.

Figure. 2. Normal ($\Delta V_{normal}$) and inverted ($\Delta V_{inverted}$) Hanle effects on (a) CoFe/MgO/Si and (b) CoFe/MgO/Ge contacts under $M \perp B$ (closed circles) and $M//B$ (open circles) measurements, respectively, at an applied current ($I$) of -500 µA (spin injection condition) for various temperatures. All curves are normalized by the voltage difference between the minimum value of the normal Hanle curve and the saturation value of the inverted Hanle curve in the region *(i)* (see figure. 1).

Figure. 3. (a) Interfacial depolarization effect ($\left|\Delta V_{inverted} / \Delta V_{normal}\right|$) and (b) effective spin lifetime ($\tau_{eff}$) as a function of the temperature ($T$).

Figure. 4. (a) Calculated normal ($M \perp B$, wine symbol) and inverted ($M//B$, brown symbols) Hanle curves. The spin lifetime ($\tau_{sf}$, blue symbols) was varied from 0.50 ns to 2.00 ns at a fixed value of $1/\omega_L^{ms}$ of about 0.33 ns, corresponding to a $B^{ms}$ value of 0.3 kOe. The ideal Hanle curves (blue symbols) without $B_L^{ms}$ are also presented for comparison. (d) Calculated the $R_{ISD}$ as a function of $\tau_{sf}$ with the four different $1/\omega_L^{ms}$ values of about 0.10, 0.33, 1.00, and $\infty$ ns, corresponding to $B_L^{ms}$ values of about 1.0, 0.3, 0.1, and 0.0 kOe, respectively. (e) $\tau_{eff}$ vs. $\tau_{sf}$ plot for different values of $1/\omega_L^{ms}$. It is noted that the $\tau_{sf}$ is the input value for calculation and the $\tau_{eff}$ is the extracted value

from the calculated normal Hanle curve (see figures. 4(a)-(c)).

Figure. 5. Normal ($\Delta V_{normal}$) and inverted ($\Delta V_{inverted}$) Hanle effects on (a) CoFe/MgO/Si and (b) CoFe/MgO/Ge contacts under $M \perp B$ (closed circles) and $M//B$ (open circles) measurements, respectively, with various reverse bias currents ($I < 0$, spin injection condition) at 5 K. All curves are normalized by the voltage difference between the minimum value of the normal Hanle curve and the saturation value of the inverted Hanle curve in the region *(i)* (see figure. 1).

Figure. 6 (a) Interfacial depolarization effect ($\left| \Delta V_{inverted} / \Delta V_{normal} \right|$) and (b) effective spin lifetime ($\tau_{eff}$) as a function of the reverse bias current ($I<0$, spin injection) at 5 K and 300 K.

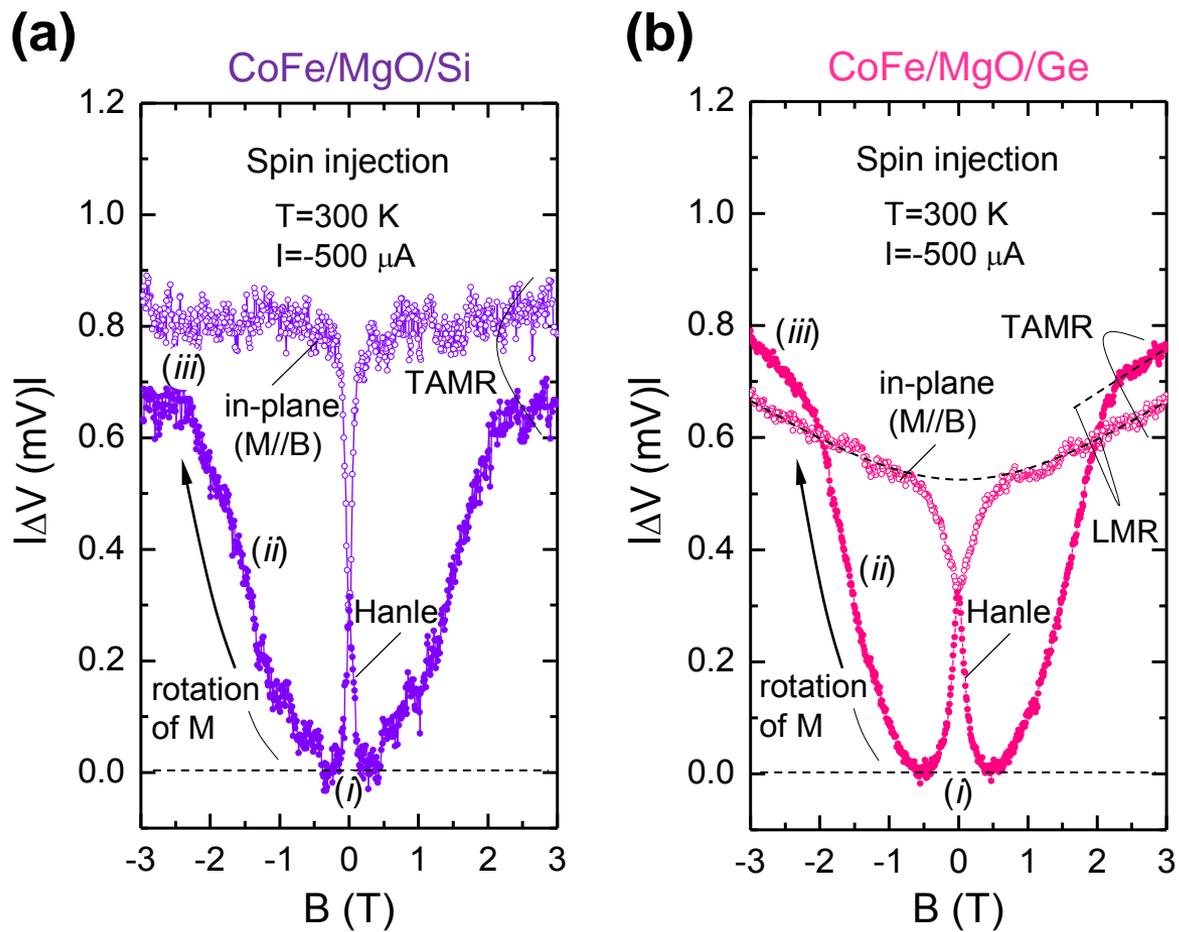

Fig. 1

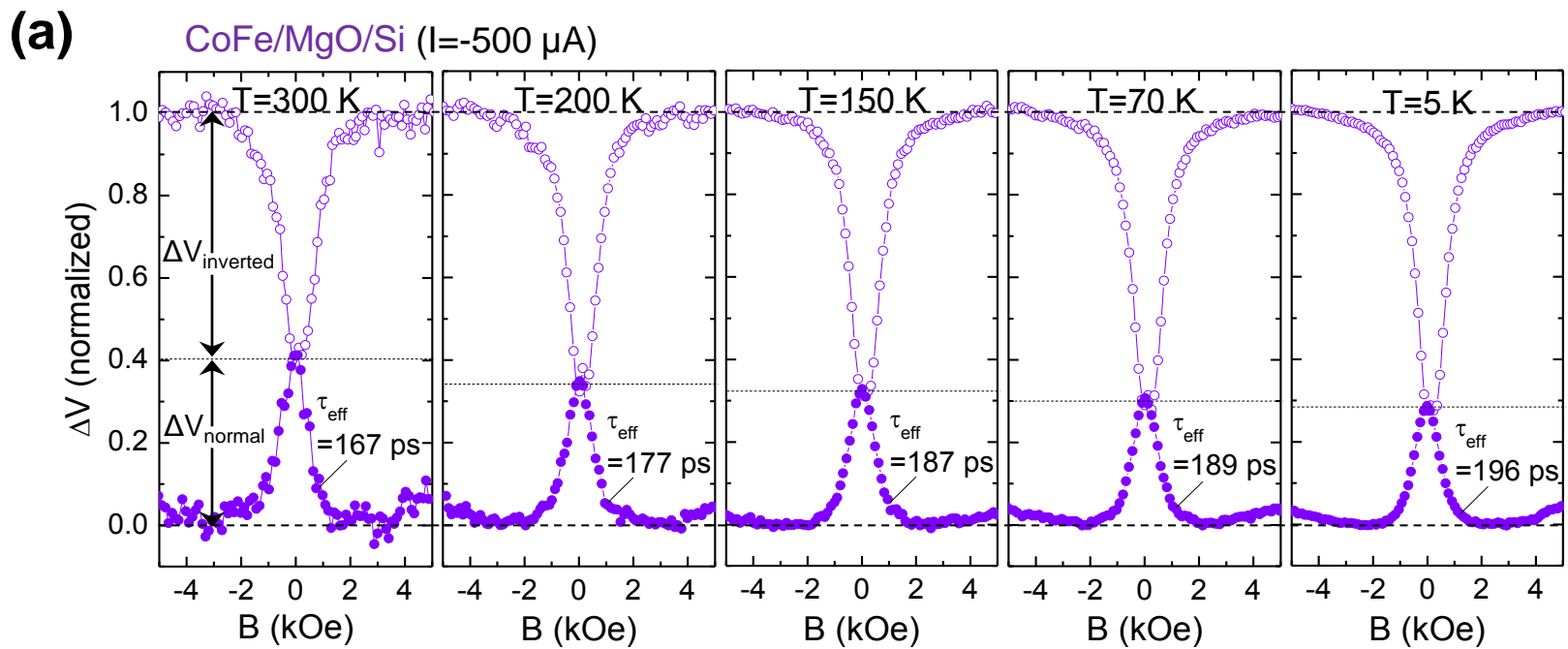
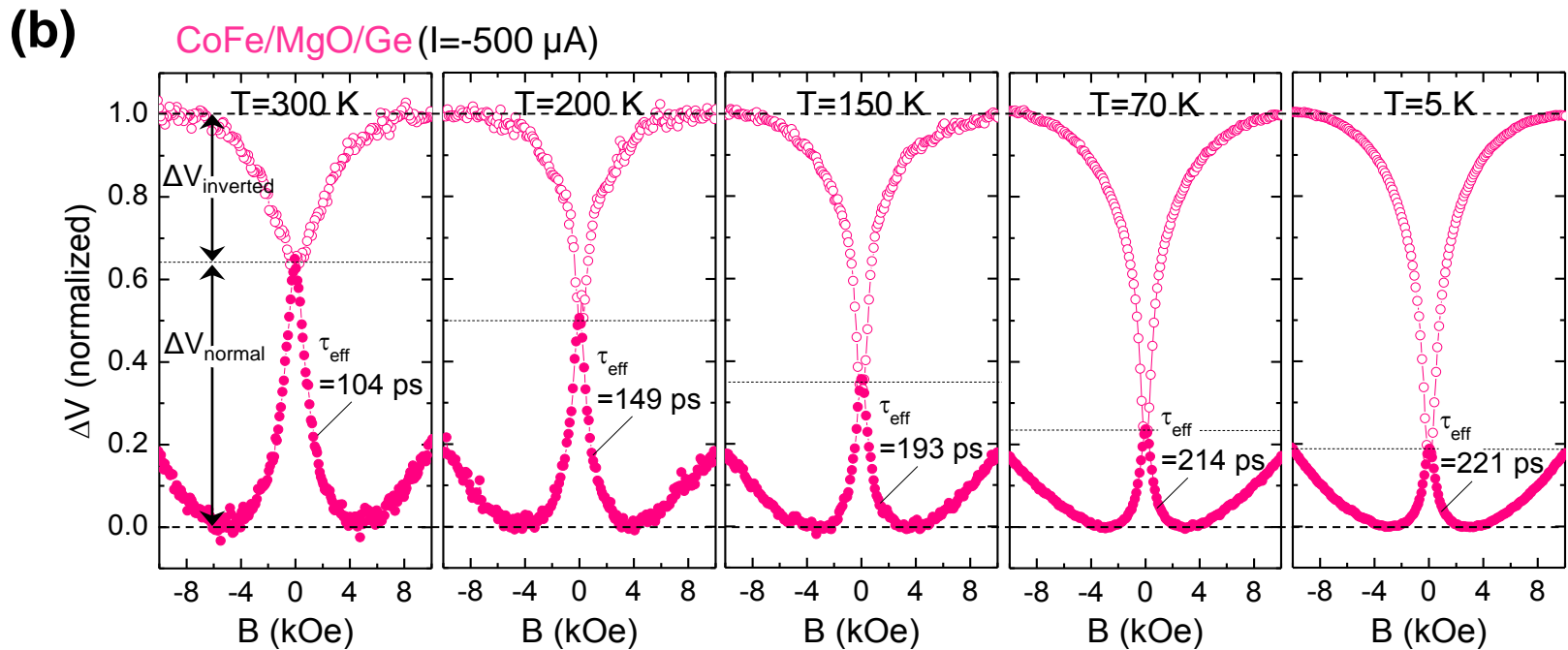

Fig. 2

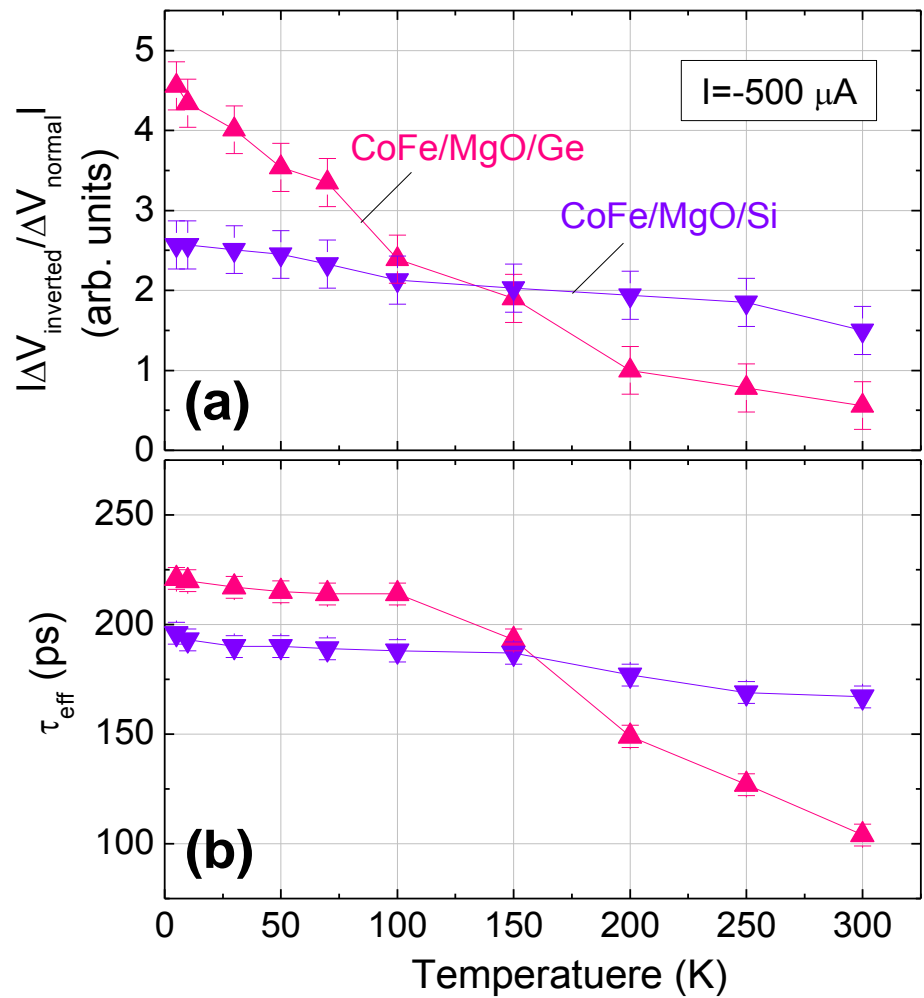

Fig. 3

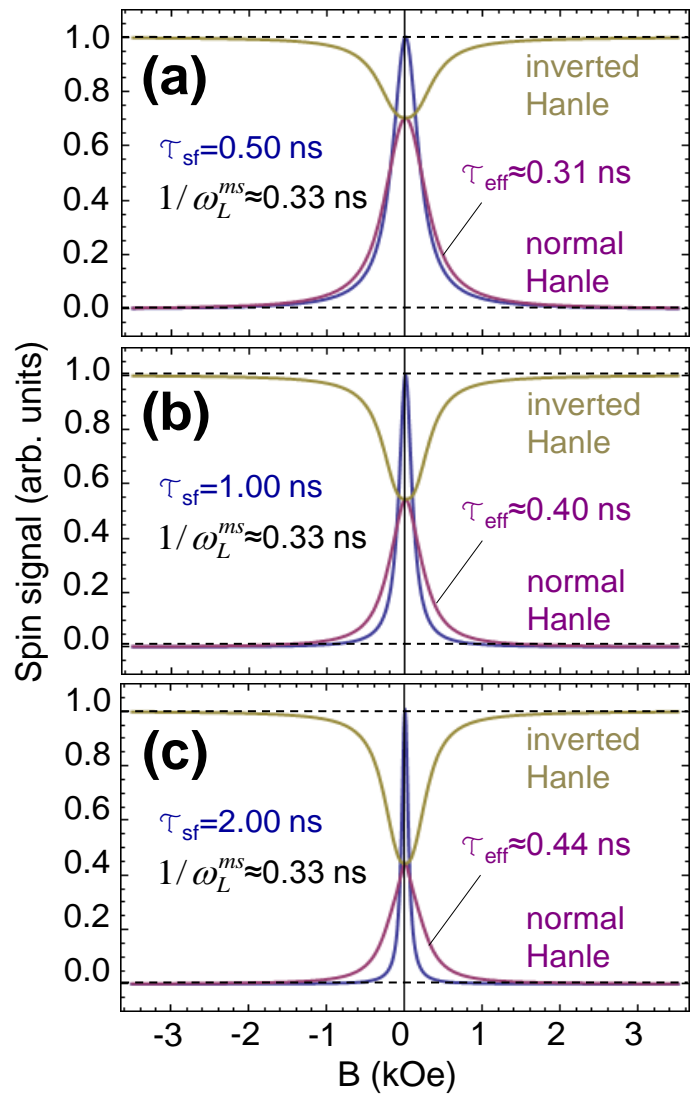
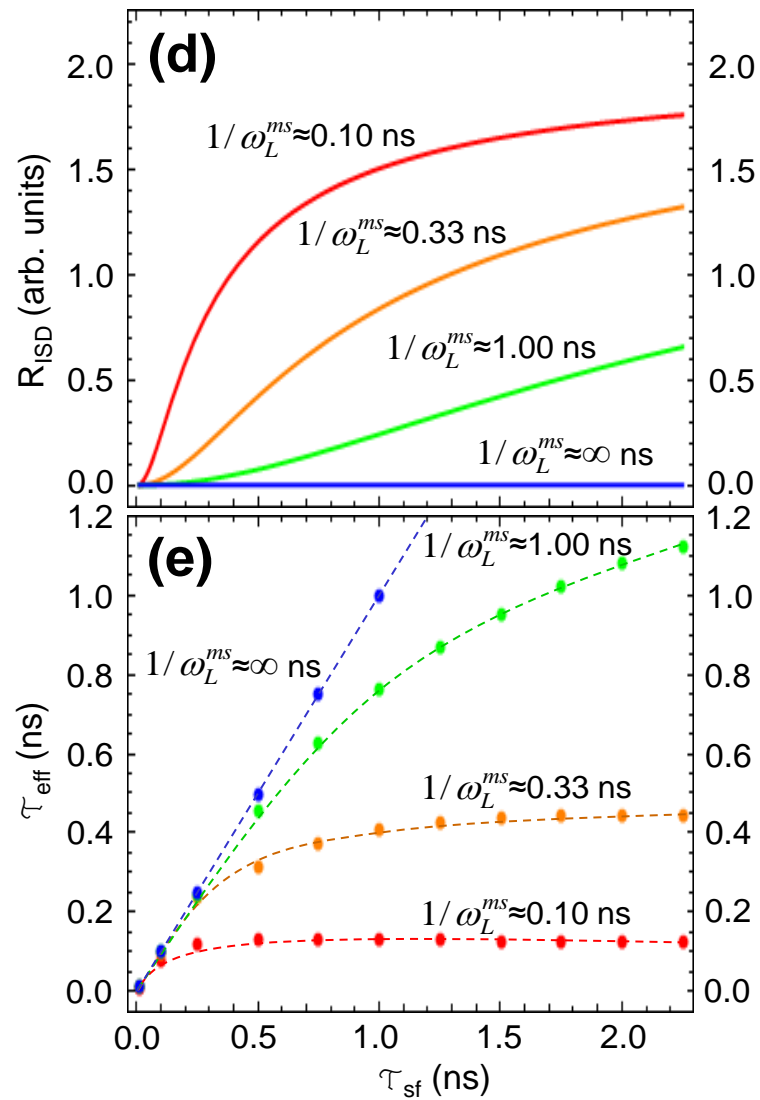

Fig. 4

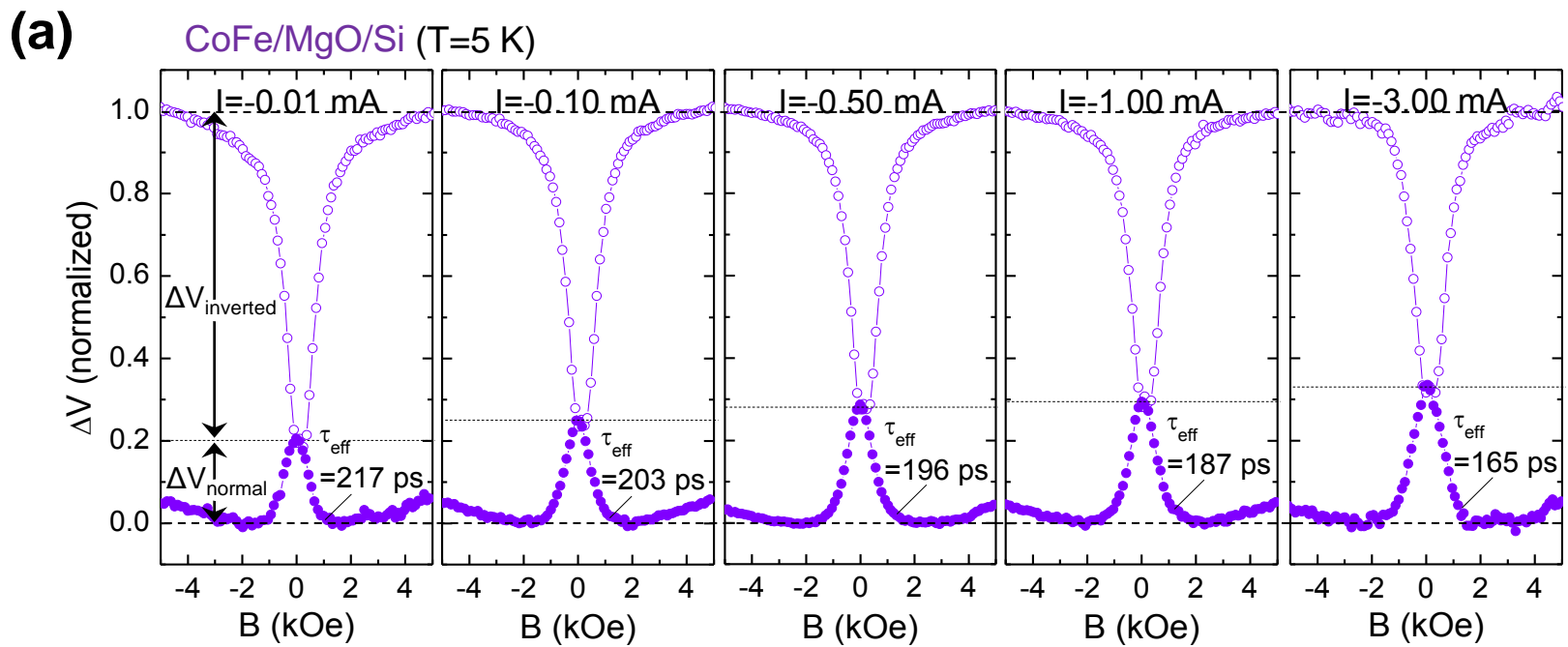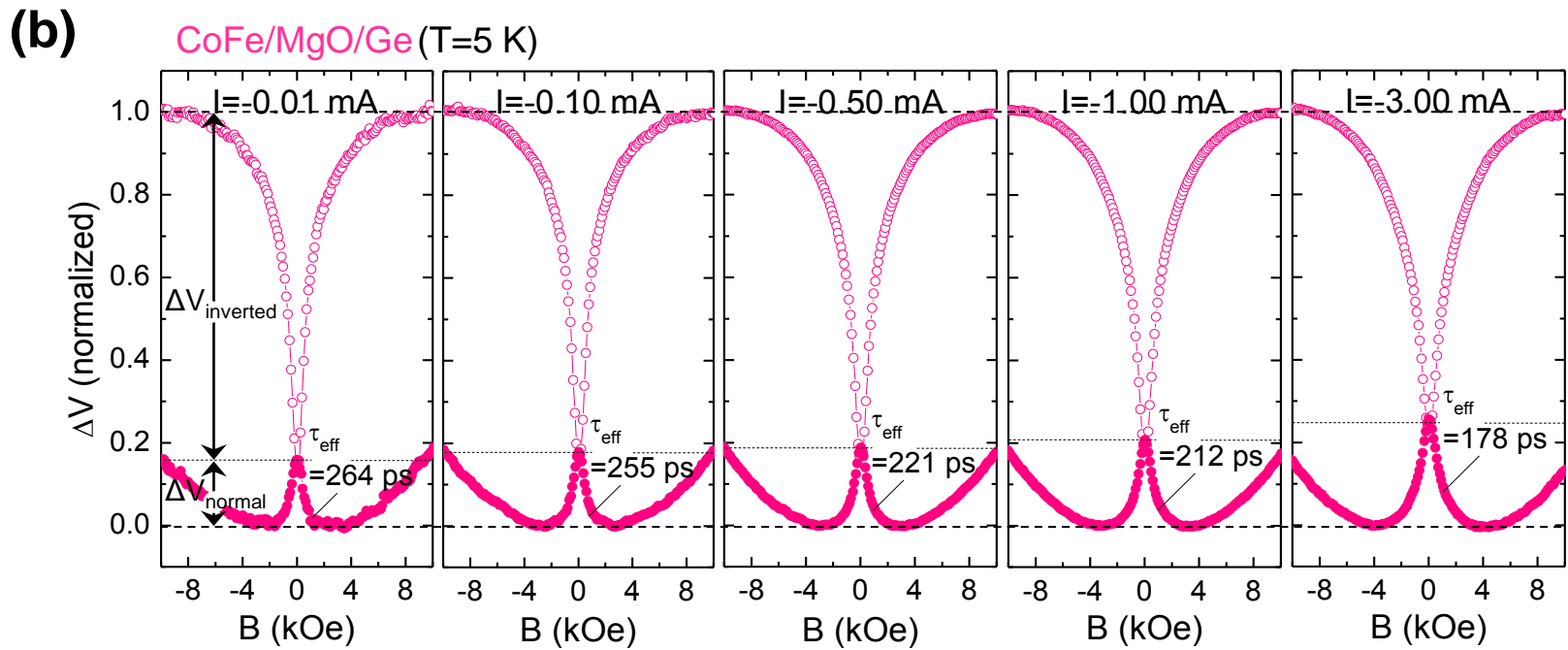

Fig. 5

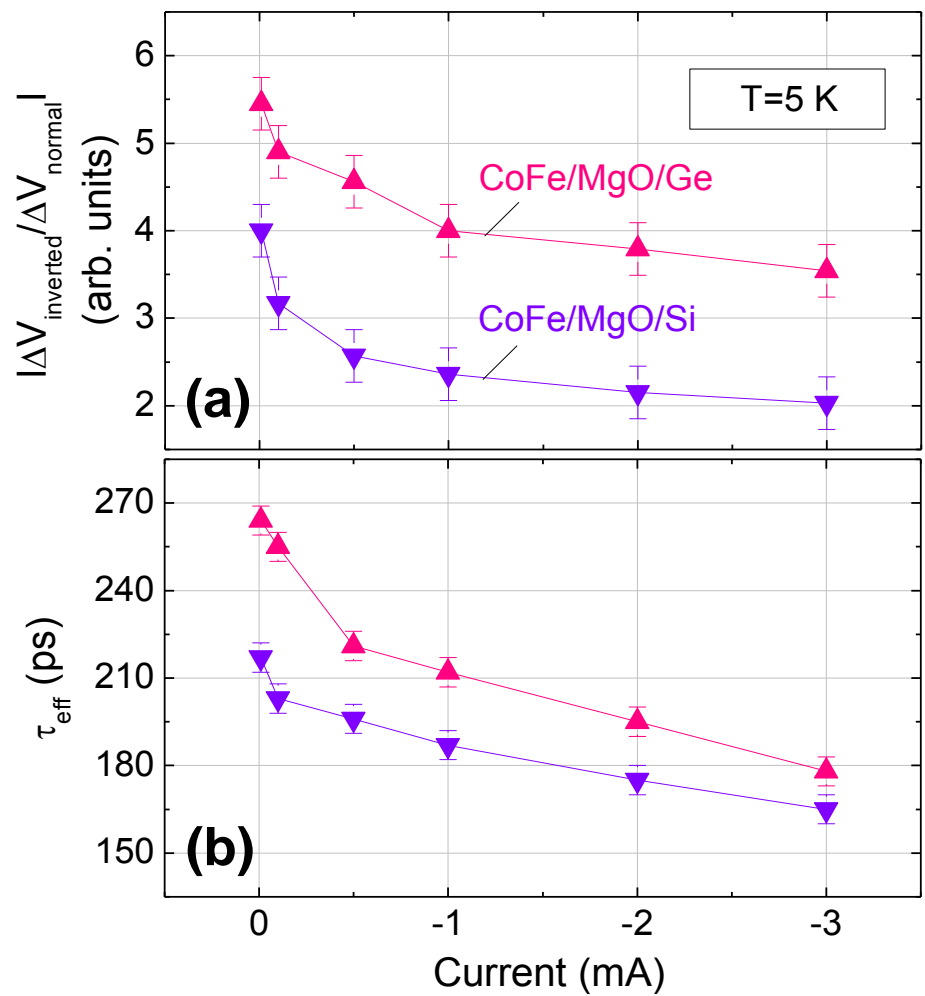

Fig. 6